\documentclass[12pt]{iopart}

\usepackage{graphicx}
\usepackage{amsfonts}

\begin{document}

\title[FT series for the figure eight solution]
{Fourier-Taylor series for the figure eight solution of the three body problem}
\author{Heinz-J\"urgen Schmidt$^1$ and Thomas Br\"ocker$^1$
}
\address{$^1$Department of Physics, University of Osnabr\"uck,
 D - 49069 Osnabr\"uck, Germany}

\ead{hschmidt@uos.de}

\begin{abstract}
We provide an analytical approximation of a periodic solution of the
three body problem in celestial mechanics, the so-called figure
eight solution, discovered $1993$ by C. Moore. This approximation
has the form of a Fourier series whose components are in turn Taylor
series w.~r.~t.~some parameter. The method is first illustrated by
application to two other problems, (1) the problem of oscillations
of a particle in a cubic potential that has a well-known analytic
solution in terms of elliptic functions and (2) periodic solutions
and corresponding eigenvalues of a generalized Mathieu equation that
cannot be solved analytically. When applied to the three body
problem it turns out that the Fourier-Taylor series, evaluated up to
30th order, represents un-physical solutions except for a particular
value of the series parameter. For this value the series
approximates the numerical solution known from the literature up to
a relative error of $1.6\times 10^{-3}$.
\end{abstract}

%

%
%
%
%


\section{Introduction}\label{sec:I}
Analytical approximations provide an access to problems that cannot
be solved analytically and have some advantages in comparison to
purely numerical solutions. Let us consider periodic solutions
${\mathbf x}(t)$ of an $n$-dimensional differential equation of the
form
\begin{equation}\label{I1}
\dot{\mathbf x}={\mathbf f}({\mathbf x})
\;.
\end{equation}
An obvious attempt to solve (\ref{I1}) is by means of Fourier series for ${\mathbf x}(t)$.
However, if ${\mathbf f}$ is non-linear one encounters the problem to evaluate the
r.~h.~s.~of (\ref{I1}). Already the quadratic term of a Taylor expansion of ${\mathbf f}({\mathbf x})$
involves the convolution of two infinite series of Fourier coefficients. Fortunately, in physical
problems it often happens that the Fourier coefficients of the pertinent quantities decay exponentially
with order. Hence one could try to truncate the infinite Fourier series thereby avoiding
the infinite convolution problem. In order to get a better approximation one has to increase the
length of the finite Fourier series stepwise and one would like to have an iterative solution algorithm
that uses the previous results for the coefficients at each
step. The method of Fourier-Taylor (FT) series is a systematic approach that satisfies these requirements.
Moreover, it solves another problem, namely that the period of the oscillation that has to be used in
the Fourier series ansatz is often unknown and can only be numerically calculated.\\
For the above problem the general FT series would read
\begin{eqnarray}\label{I2a}
{\mathbf x}(t)&=&\sum_{n\in{\mathbb Z}}\sum_{m=|n|}^\infty \,{\mathbf A}_{nm}\,\lambda^m \, \exp (i\, n\, \omega\, t) \\
\label{I2b}
\omega &=&\sum_{m=0}^\infty\, \omega_m \lambda^m
\;.
\end{eqnarray}
Hence each Fourier component of ${\mathbf x}(t)$ of order $n$ is a Taylor series w.~r.~t.~a parameter $\lambda$
that starts with a term proportional to $\lambda^n$. Put differently, ${\mathbf x}(t)$ is written as a Taylor series
w.~r.~t.~$\lambda$ such that each Taylor coefficient of $\lambda^m$ is a finite Fourier series at most of order $m$.
Hence, if the r.~h.~s.~of (\ref{I1}) is expanded into a Taylor series w.~r.~t.~$\lambda$ we only encounter convolutions of finite order.
A comparison of the coefficients of $\lambda^m \, \exp (i\, n\, \omega\, t) $, $n=0,\ldots,m$ yields
a sequence of equations that can be used to iteratively determine the unknowns ${\mathbf A}_{nm}$ and $\omega_m$.
This comparison is performed by expanding $\omega$ only at the l.~h.~s.~of (\ref{I1}); at the r.~h.~s.~of (\ref{I1})
the $\omega$ occurring in $\exp (i\, n\, \omega\, t) $ is left intact.\\
Up to now, $\lambda$ appeared as a purely formal parameter that is only used for book-keeping. However, this turns out to be
problematic.
Nothing prevents one to use, say, $\sinh(\lambda)$ instead of $\lambda$ as the series parameter. This transformation would also
influence the unknown coefficients and hence these coefficients {\it a priori} cannot be uniquely determined.
Already counting the number of equations and unknowns indicates this kind of under-determination.
The way out of this problem
is a ``concretization" of $\lambda$, i.~e.~, $\lambda$ has to be chosen as a parameter with a concrete meaning,
e.~g.~the amplitude of the oscillation. This will modify the FT ansatz (\ref{I2a}) and, hopefully,
give unique solutions for ${\mathbf A}_{mn}$ and $\omega_m$. Nevertheless, the first coefficients are often
ambiguous and some choice has to be made, see the examples of the following sections.
This is due to the effect that the very first equations are quadratic or of higher order and become linear ones only after
a few steps.\\

The method of FT series mostly cannot be successfully applied without carefully considering the proper choice of the
variables and of the differential equation. Often one has some information about the solutions that can be
used to simplify the ansatz (\ref{I2a}) and (\ref{I2b}). This is also exemplified in the following sections.
Generally speaking, the method of FT series can be viewed as a generalization of the method of linearizing equations
like (\ref{I1})
and, similarly, its quality of approximation depends on the size of the parameter $\lambda$ and of the maximal order of
truncation. Since the basic idea is very simple we would not be surprised to learn that the method has already been used
before, although we do not know of any application in the literature. One reason for this may be that the use of high truncation orders
requires computer algebra software that has been only available during the last decades.
To our best knowledge, the method has first been
used in a lecture of one of the authors on non-linear wave equations \cite{HJS03}.\\

The paper is organized as follows.
In order to illustrate the concrete application of the FT series method and to demonstrate its
performance for a simple but non-trivial example we consider, in section \ref{sec:C}, the oscillations of a particle in a cubic
potential (an-harmonic oscillator). Since the analytical solution
of this problem is well-known it can be utilized to test the quality of approximation provided by the method.
We expand the FT series up to $24$th order and obtain a precision of $5\times 10^{-15}$ for a medium amplitude that is  $1/6$
of the maximal value obtained for the aperiodic limit case.\\
Secondly, in section \ref{sec:M} we consider periodic solutions of a generalized Mathieu equation of Hill's type. Possible
physical applications include the Schr\"odinger equation of a particle in a periodic potential and diffraction
of electromagnetic waves in a periodic grating.
In contrast to
the an-harmonic oscillator this is a linear problem such that the Fourier series ansatz leads to an infinite matrix problem.
Periodic solutions exist only for certain values of a parameter $a$ usually called ``characteristic values" of the differential equation.
We will refer to these as ``eigenvalues" in accordance with the nomenclature commonly used in physics.
Interestingly, the FT series leads to some kind of perturbation series for the eigenvalues and eigenfunctions without using
the methods of perturbation theory. For a similar approach to holographic diffraction see \cite{SIV14}.\\
As the main application we consider, in section \ref{sec:E}, the periodic solution of the three body problem
that has been found two decades ago \cite{M93} and mathematically investigated in \cite{CM00}, but where an analytical solution is not available.
Three equal masses perform a periodic motion following the same orbit that has the
form of the figure eight with a constant relative time delay of $T/3$ (``choreography"). Without loss of generality the period $T$
can be chosen as $T=2\pi$. Usually the FT series method yields
a one-parameter family of periodic solutions as in the example considered in section \ref{sec:C}. For the figure eight
solution we also obtain such a family but it will be completely un-physical due to its violation of angular momentum
conservation, except for a particular value $\lambda_1\approx  2.07552$ of the parameter $\lambda$. For this value
the figure eight solution numerically calculated in \cite{CM00} is reproduced with a relative error
of less than $1.6\times 10^{-3}$ if we expand the FT series up to $30$th order.
We close with a summary and outlook.\\
The detailed results of the FT series method for the three examples are too complicated to be presented in a paper and will be
given in three MATHEMATICA 10.1 notebooks that can be downloaded from \cite{S15}. Further applications of the FT method
will be given elsewhere \cite{SSHL15}.

\section{FT series for oscillations in a cubic potential}
\label{sec:C}
\begin{figure}
\begin{center}
\includegraphics[clip=on,width=90mm,angle=0]{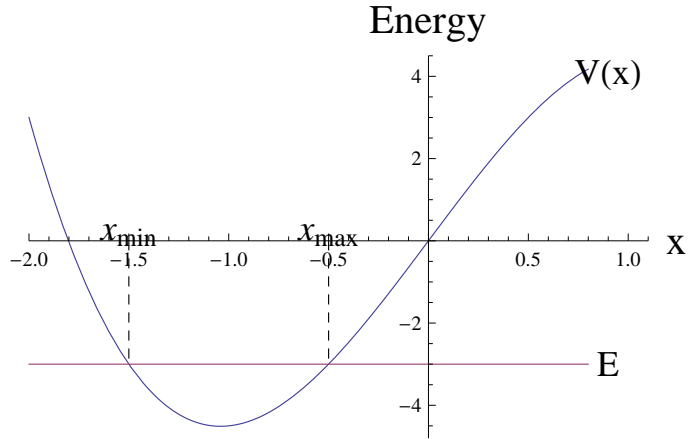}
\end{center}
\caption{Illustration of the potential $V(x)$ according to (\ref{C1}) and the values of $a=1$ and $\lambda=1/2$ corresponding to a
total energy of $E=-3$.
The particle performs non-linear oscillations between the positions $x_{min}=-1.5$ and $x_{max}=-0.5$ .
}
\label{FIGPW}
\end{figure}

We consider the oscillations of a particle
with unit mass in a potential
\begin{equation}\label{C1}
V(x)=
2 x \left(3 a^2+\lambda ^2\right)-2 x^3
\;,
\end{equation}
and total energy
\begin{equation}\label{C2}
E=-4 a (a-\lambda ) (a+\lambda )
\;,
\end{equation}
where $a$ and $\lambda$ are positive parameters. The potential is
shown in figure \ref{FIGPW} for the choice of $a=1$ and $\lambda=1/2$.

\begin{figure}
\begin{center}
\includegraphics[clip=on,width=90mm,angle=0]{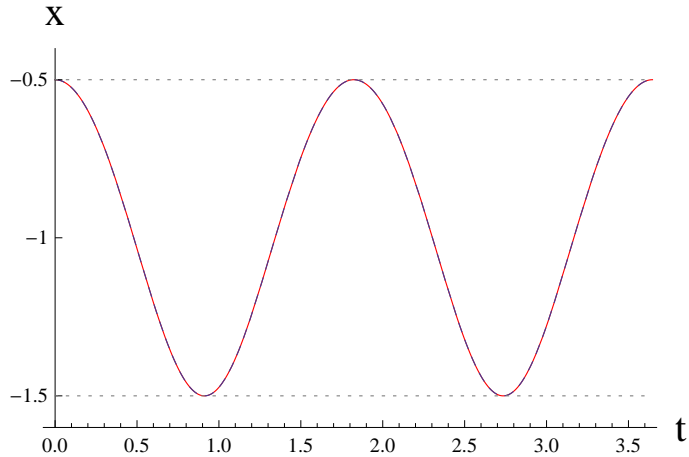}
\end{center}
\caption{Oscillation of a particle in the cubic potential of figure \ref{FIGPW}
according to the analytical solution (\ref{C12})
(dashed curve) and the FT series of 24th order (red curve).
}
\label{FIGA1}
\end{figure}

\begin{figure}
\begin{center}
\includegraphics[clip=on,width=90mm,angle=0]{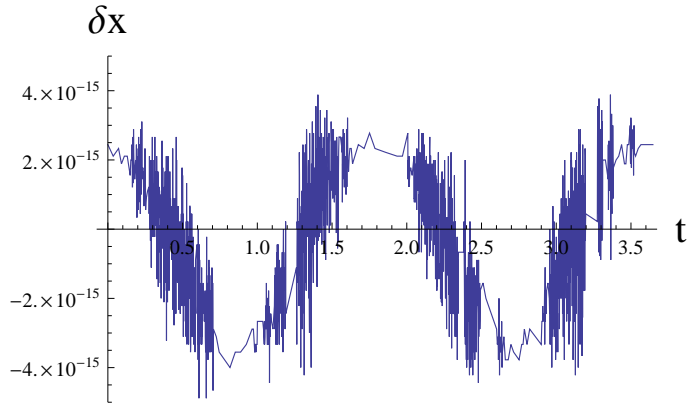}
\end{center}
\caption{Difference between the two solutions displayed in figure \ref{FIGA1}.
}
\label{FIGA2}
\end{figure}

Energy conservation yields the differential equation
\begin{eqnarray}\label{C3}
\left(\frac{dx}{dt}\right)^2
&=&
2(E-V(x))=4 (x-2 a) (x-\lambda +a ) (x+\lambda +a )
\\
\label{C3a}
&\equiv& 4x^3-g_2 x-g_3\equiv P(x)
\;.
\end{eqnarray}
Note that this differential equation differs from the form given in (\ref{I1}),
but it has the advantage that its r.~h.~s.~is a polynomial in $x$ and hence has a very simple
Taylor expansion.\\
It is easily shown that $P(x)$ is positive in the range
$x_{min}\equiv -a-\lambda < x < -a+\lambda \equiv x_{max}$
and hence the particle performs oscillations between the positions
$x_{min}$ and $x_{max}$. In the limit $\lambda\rightarrow 0$ we obtain
the asymptotic solution
\begin{eqnarray}\label{C4a}
x(t)&=& -a +\lambda\,\cos\,\omega t+ {\mathcal O}(\lambda^2),\\
\label{C4b}
\omega&=& 2\sqrt{3\, a}\,+\, {\mathcal O}(\lambda^2)
\;.
\end{eqnarray}
Here and in what follows we chose the point of zero time such that the maximal amplitude is assumed at $t=0$.\\

For the general solution we make the following FT series ansatz:
\begin{eqnarray}\label{C5a}
x(t)&=&\sum_{n=0}^\infty\,\sum_{m=n}^\infty\,A_{n\,m}\lambda^m\,\cos\,n\,\omega\, t\,\\
\label{C5b}
\omega&=&\sum_{m=0}^\infty\,\omega_m\,\lambda^m
\;.
\end{eqnarray}
The FT series contains only $\cos$-terms since $x(t)$ is an even function for the
above choice of the zero time point. This ansatz is inserted into
the differential equation (\ref{C3}). Both sides of (\ref{C3}) must have the same
coefficients of the terms $\lambda^m\,\cos\,n\,\omega\,t$. This yields a sequence
of equations that are used to determine the unknowns $\omega_m$ and $A_{n\,m}$
to arbitrary order that is only limited by computer resources.
It is advisable to do the first steps ``by hand".

The coefficient of $\lambda^0$ of the l.~h.~s.~of (\ref{C3}) vanishes,
whereas the r.~h.~s.~of (\ref{C3}) gives $4 (A_{0\,0}-2a)(A_{0\,0}+a)^2$.
We choose the solution $A_{0\,0}=-a$ in view of (\ref{C4a}). The coefficient of $\lambda^1$
vanishes at both sides of (\ref{C3}). Comparison of coefficients of
$\lambda^2\,\cos\,n\,\omega\,t$ for $n=0,1,2$ yields the following three equations:
\begin{eqnarray}\label{C6a}
0&=& 12 a (2 A_{0\,1}^2+A_{1\,1}^2-2)+A_{1\,1}^2\,\omega_0^2\;,\\
\label{C6b}
0&=&a\,A_{0\,1}\,A_{1\,1}\;,\\
\label{C6c}
0&=&A_{1\,1}\,(\omega_0^2-12 a)
\;.
\end{eqnarray}
We choose the solution $A_{0\,1}=0,\; A_{1\,1}=1,\;\omega_0=2\sqrt{3\, a}$
in accordance with (\ref{C4a}) and (\ref{C4b}). After these choices the comparison of coefficients
of orders $m\ge 3$ yields unique results at least up to order 24. It turns out that the coefficients of the powers
$\lambda^m$ are only non-zero if $m$ increases in steps of $2$ in accordance with the above solution $A_{0\,1}=0$.
The detailed results are too complicated to be presented here and can be found in \cite{S15}.
We only give the next order corrections to (\ref{C4a}) and (\ref{C4b}):
\begin{eqnarray}\label{C7a}
x(t)&=& -a +\frac{\lambda^2}{12 a}+\left(\lambda-\frac{\lambda^3}{192 a^2}\right)\,\cos\,\omega t
-\frac{\lambda^2}{12 a}\cos 2\,\omega\,t+ {\mathcal O}(\lambda^4),\\
\label{C7b}
\omega&=& 2\sqrt{3\, a}-\frac{\lambda^2}{8 \sqrt{3} a^{3/2}}\,+\, {\mathcal O}(\lambda^4)
\;.
\end{eqnarray}

Finally we compare our FT series solution of order 24 with the analytical solution for the special values
$a=1$ and $\lambda=1/2$. For the frequency $\omega$ and the corresponding oscillation period $T$
we obtain
\begin{equation}\label{C8}
\omega\approx 3.4458763864722597,\quad T=\frac{2\pi}{\omega}\approx 1.8233925429960192
\;.
\end{equation}
The period can also be obtained by the following integral
\begin{eqnarray}\label{C9a}
T&=&2\int_{-\frac{3}{2}}^{-\frac{1}{2}} \frac{1}{\sqrt{4 (x-2)
   \left(x+\frac{1}{2}\right) \left(x+\frac{3}{2}\right)}} \, dx\\
   \label{C9b}
&=&2 \sqrt{\frac{2}{7}}\; K\left(\frac{2}{7}\right)=1.823392542996019014655\ldots
   \;,
\end{eqnarray}
where $K(m)$ denotes the complete elliptic integral of first kind, see
\cite{AS72} Ch.~17. We note that the relative error is of order $10^{-16}$.

Upon defining
\begin{equation}\label{C10}
u_1\equiv \int_{\infty }^{-\frac{1}{2}} \frac{1}{\sqrt{4 (x-2)
   \left(x+\frac{1}{2}\right) \left(x+\frac{3}{2}\right)}} \, dx
   = -\sqrt{\frac{2}{7}} K\left(\frac{2}{7}\right)+i \sqrt{\frac{2}{7}}
   K\left(\frac{5}{7}\right)
   \;,
\end{equation}
and evaluating the coefficients of $P(x)$
\begin{equation}\label{C11}
g_2= 4 \left(3 a^2+\lambda ^2\right)=13,\;g_3= 8 \left(a^3-a \lambda ^2\right)=6
\;,
\end{equation}
we obtain after some standard transformations
\begin{equation}\label{C12}
x(t)={\mathcal P}(t-u_1;g_2,g_3)
\;,
\end{equation}
where ${\mathcal P}(z;g_2,g_3)$ denotes the Weierstrass elliptic function, see \cite{AS72}, Ch.~18.
The coincidence of this analytical result with the above FT series of order $24$ is very close, see the figures \ref{FIGA1}
and \ref{FIGA2}. Of course, the quality of the approximation would decrease when approaching the aperiodic solution corresponding to
$\lambda=3a$. We will not closer investigate this problem since our intention was only to provide
an example where the method of FT series gives reasonable results.

\section{FT series for the generalized Mathieu equation}
\label{sec:M}
\begin{figure}
\begin{center}
\includegraphics[clip=on,width=120mm,angle=0]{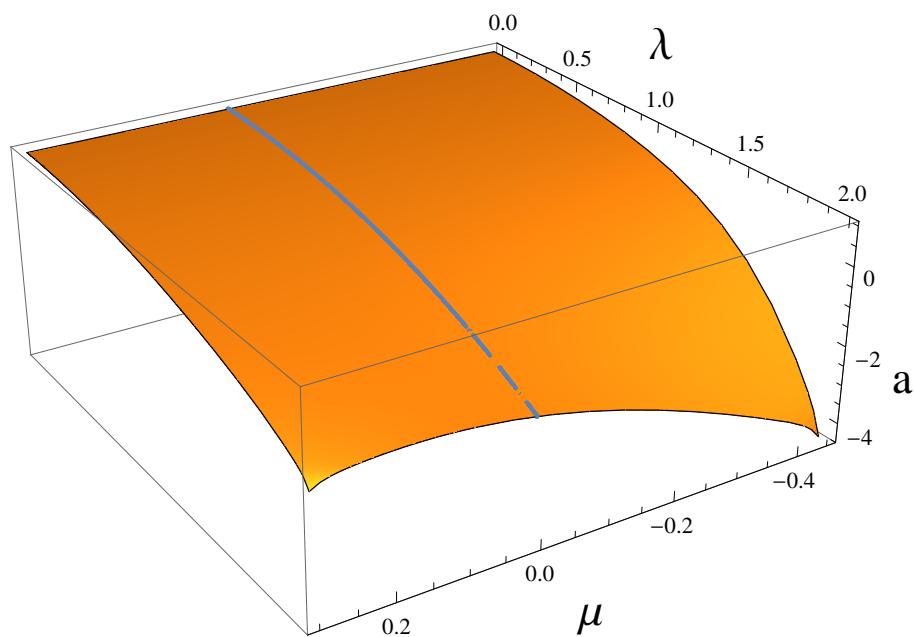}
\end{center}
\caption{The lowest eigenvalue $a$ for odd solutions of the generalized Mathieu equation (\ref{M1}) as a function of the parameters
$\lambda$ and $\mu$. For $\mu=0$ the lowest eigenvalue can be calculated by the MATHEMATICA command $\mbox{ MathieuCharacteristicB}[1, \lambda^2/2]$,
see the blue curve.
}
\label{FIGM1}
\end{figure}

\begin{figure}
\begin{center}
\includegraphics[clip=on,width=120mm,angle=0]{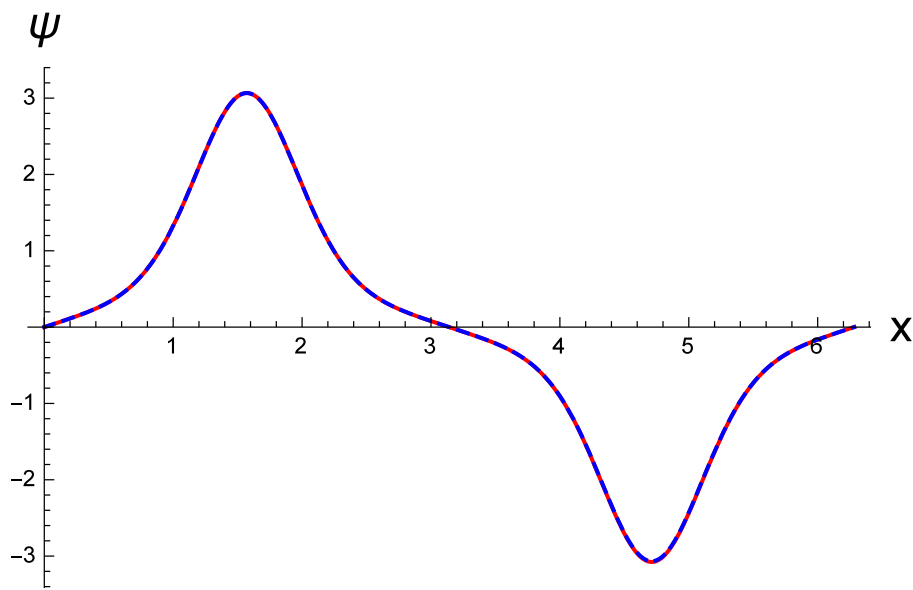}
\end{center}
\caption{A special periodic solution $\psi(x)$ of the generalized Mathieu equation (\ref{M1})
with $\lambda=2,\,\mu=-0.3$ according to the FT series (blue dashed curve) and the corresponding
numerical solution (red curve).
}
\label{FIGM2}
\end{figure}

\begin{figure}
\begin{center}
\includegraphics[clip=on,width=120mm,angle=0]{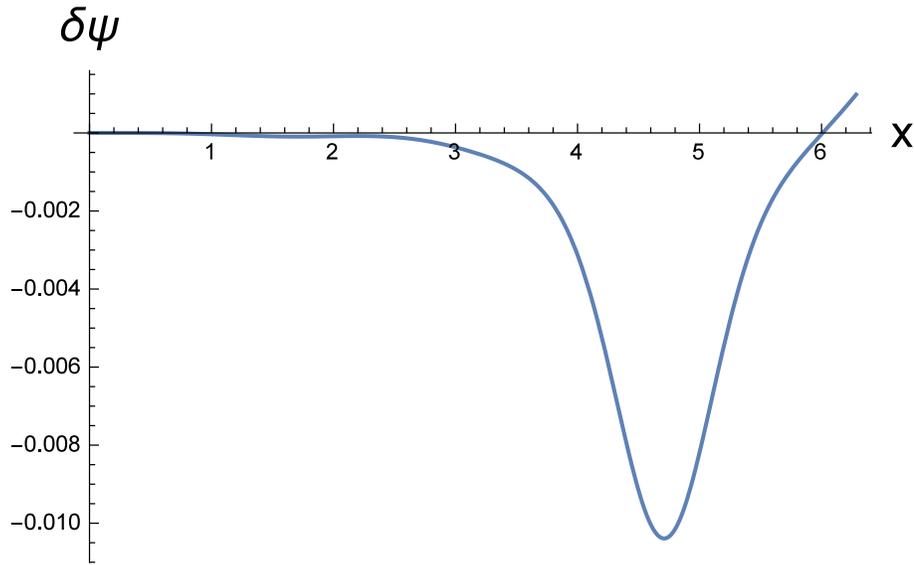}
\end{center}
\caption{The difference $\delta\,\psi$ between the two solutions show in figure \ref{FIGM2}.
}
\label{FIGM3}
\end{figure}

As a further application of the FT series approximation we consider a linear differential equation
that has periodic solutions with period (or wavelength) $T=2\pi$ for the eigenvalues $a$,
namely a generalized Mathieu equation of the form
\begin{equation}\label{M1}
\frac{d^2}{dx^2}\psi = -a +\lambda^2\,\cos(2x)+\lambda^4\,\mu\,\cos(4x)
\;.
\end{equation}
We have called the independent variable $x$ and the unknown function $\psi(x)$ because of the association with the
one-dimensional Schr\"odinger equation for a particle in a periodic potential.
For $\mu=0$ this is the ordinary Mathieu equation \cite{AS72}, Ch.~20, with $\lambda^2=2\,q$
and its periodic solutions including the corresponding eigenvalues
are known special functions that be calculated by using computer-algebraic commands. However, the corresponding eigenvalues and solutions of the
generalized Mathieu equation could only be numerically calculated, e.~g., by truncating an infinite-dimensional matrix problem.\\
Since we only want to illustrate the application of the FT method it will suffice to consider odd solutions corresponding
to the lowest eigenvalue of (\ref{M1}). For $\lambda=0$ this is the solution $\psi(x)=\lambda\,\sin x$ with the eigenvalue $a=1$,
where we have set the amplitude to $\lambda$ according to the FT ansatz (\ref{I2a}). In the same way the pre-factors
$\lambda^2$ and $\lambda^4$ in (\ref{M1}) are dictated by the FT ansatz, whereas $\mu$ is a free real parameter.
The FT ansatz for odd solutions of (\ref{M1}) is hence chosen as
\begin{eqnarray}\label{M2a}
\psi(x)&=&\lambda\,\sin(x)+\sum_{n=3,5,\ldots}\sum_{m=n,n+2,\ldots}\,\Psi_{n,m}\,\lambda^m\,\sin(n\,x),\\
\label{M2b}
a&=&\sum_{m=0,2,4,\ldots}\,a_m\,\lambda^m
\;.
\end{eqnarray}
The coefficients $\Psi_{n,m}$ and $a_m$ can be calculated relatively rapidly due to the linearity of the problem.
Hence we have chosen a maximal truncation order of $n=99$. The results are again too complicated to be
presented here and can be found in \cite{S15}. We only show the first few
terms of the FT series for the lowest eigenvalue:
\begin{equation}\label{M3}
a=1-\frac{\lambda ^2}{2}-\frac{\lambda ^4}{32}+\frac{\lambda ^6 }{512} (32 \mu   +1)+
\frac{\lambda ^8}{24576} \left(-1024 \mu ^2-64 \mu   -1\right)+{\mathcal O}(\lambda^{10})
\;.
\end{equation}
For $\mu=0$ the these coefficients coincide with those given in \cite{AS72} 20.2.25 if
$\lambda^2=2\,q$ is taken into account.
Figure \ref{FIGM1} shows the $\lambda,\,\mu$-dependence of $a$ for a parameter region where we expect
convergence of the FT series. As an example of a periodic solution we have chosen the values
$\lambda=2,\,\mu=-0.3$ and calculated the corresponding FT series as well as a numerical
solution of (\ref{M1}) with the same initial conditions and eigenvalue as the FT solution.
The result is shown in the figures \ref{FIGM2} and \ref{FIGM3} and demonstrates that the
FT method satisfactorily works for this example. We conjecture that the periodic solution
is unstable in this case and hence the distance to the FT approximation increases with $x$.

\section{FT series for the figure eight solution}
\label{sec:E}
\begin{figure}
\begin{center}
\includegraphics[clip=on,width=120mm,angle=0]{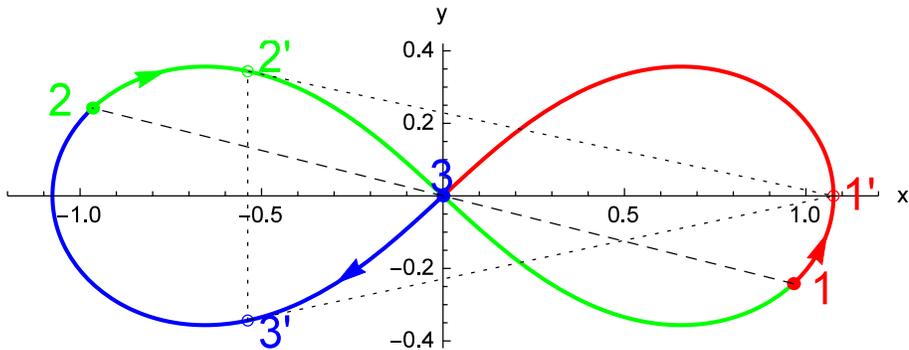}
\end{center}
\caption{Illustration of the periodic figure eight solution of the three body problem. Without loss of generality the period
is chosen as $T=2\pi$. At time $t=0$ the three
masses start at the positions marked by $1,2,3$ that are joined by the dashed line. They reach the positions marked by $1',2',3'$
joined by the dotted triangle at time $t=\frac{2\pi}{12}$. The motion in the time interval $[0,\frac{2\pi}{12}]$ completely
determines the rest of the motion that can be obtained by suitable reflections and permutations of particles. At the time
$t=\frac{2\pi}{3}$ the three masses assume their positions at $t=0$ up to a cyclic permutation (``choreography").
}
\label{FIGE1}
\end{figure}

The figure eight solution, see figure \ref{FIGE1}, is a stable plane periodic solution of the three body problem
with equal masses that has been found numerically
first in \cite{M93} and later been established by a rigorous existence proof \cite{CM00}. It has attracted considerable interest
\cite{GAM14}, \cite{S14}, \cite{SD11}, \cite{FFO03}, and
has also been generalized to larger numbers of bodies \cite{CGMS02}. One remarkable feature of it is that all three bodies follow the same curve
but with a constant phase shift (``choreography").
This is similar to the property of a spin wave in a finite spatial structure, see \cite{SSHL15}.
Nevertheless, an analytical solution is not known.
The Fourier components of this solution decay rapidly in order, hence a direct  approximation by an FT series seems possible at first sight.
Moreover, the choreographic character of the figure eight solution can be easily implemented into the FT series. \\

Let us first consider the equation of motion. We have three bodies with equal masses $m$, treated as point-like particles,
that interact via gravitational forces. Anticipating that we consider periodic solutions with a period $T$ we choose
$t_0=\frac{T}{2\pi}$ as the unit of time and $x_0=\left(m\,\gamma\,t_0^2\right)^{1/3}$ as the unit of length,
where $\gamma$ denotes the gravitational constant. Passing to dimensionless quantities,
denoting the position of the three bodies by
${\mathbf r}_\nu(t)=\left(\begin{array}{cc} x_\nu (t)\\y_\nu(t)\end{array}\right)\equiv  x_\nu (t)+ i\, y_\nu(t) ,\; \nu=1,2,3\;$
and time derivatives by a dot,
we obtain the following equations of motion
\begin{equation}\label{E0}
\ddot{\mathbf r}_1= \frac{{\mathbf r}_2-{\mathbf r}_1}{|{\mathbf r}_2-{\mathbf r}_1|^3}
+
\frac{{\mathbf r}_3-{\mathbf r}_1}{|{\mathbf r}_3-{\mathbf r}_1|^3}
\;,
\end{equation}
and analogously for $\ddot{\mathbf r}_2$ and $\ddot{\mathbf r}_3$ where the corresponding equations are obtained by cyclic permutations of $1,2,3$
(henceforwards denoted by cP). This equation of motion is invariant under the inhomogeneous Galilei group in two dimensions
($6$ parameters) and Kepler transformations ($1$ parameter). By a ``Kepler transformation" we mean the multiplication of all lengths
by $\mu^2$ and all times by $\mu^3$, where $\mu>0$ is some parameter. This transformation underlies the third Kepler law and is also
utilized in the above introduction of dimensionless quantities.
Utilizing Galilei invariance one chooses an inertial system such that the center of mass is always fixed to the origin:
\begin{equation}\label{E_CM1}
{\mathbf r}_1+{\mathbf r}_2+{\mathbf r}_3={\mathbf 0}
\;,
\end{equation}
and consequently
\begin{equation}\label{E_CM2}
\dot{\mathbf r}_1+\dot{\mathbf r}_1+\dot{\mathbf r}_1={\mathbf 0}
\;.
\end{equation}
Further constants of motion are the total energy and angular momentum that has the value $L=0$ for the figure eight solution.\\
The numerical figure eight solution suggests an FT series that starts with the terms
\begin{eqnarray}\label{E1a}
x_3(t)&=&A_{1,1}\,\lambda \, \sin t +\ldots\\
\label{E1b}
y_3(t)&=& B_{2,2}\,\lambda^2\, \sin 2 t +\ldots
\;.
\end{eqnarray}
However, this ansatz leads to the problem that the r.~h.~s.~of (\ref{E0}) is not differentiable at $\lambda=0$ and hence the FT series does
not exist. Attempts to weaken the strict correspondence between Fourier orders and powers of $\lambda$ avoid this problem but cause
new ones. Another problem is that the FT series usually yields a one-parameter family of periodic solutions,
as in the example of section \ref{sec:C}, but such a non-trivial family of figure eight solutions is not known.
By ``trivial" families of figure eight solutions we mean the transformations induced by the invariance group of (\ref{E0}) mentioned above.
Hence the FT series method is apparently not suited to treat the problem at hand.\\

Yet we start a new attempt by considering relative coordinates and defining
\begin{equation}\label{E2}
{\mathbf r}_{12}\equiv {\mathbf r}_{1}-{\mathbf r}_{2} = r_{12} \exp ( i \,\varphi_{12})\quad \mbox{ and cP}
\;.
\end{equation}
After some steps the new equation of motion is written as
\begin{eqnarray}\label{E3a}
\ddot{r}_{12}&=&r_{12}\,\dot{\varphi}_{12}^2-2\,r_{12}^{-2}+r_{31}^{-2}\,\cos(\varphi_{12}-\varphi_{31})
+r_{23}^{-2}\,\cos(\varphi_{12}-\varphi_{23}),\\
\label{E3b}
\ddot{\varphi}_{12}&=&r_{12}^{-1}\left( -2\,\,\dot{r}_{12}\,\dot{\varphi}_{12}\,-r_{31}^{-2}\,\sin(\varphi_{12}-\varphi_{31})
\,-r_{23}^{-2}\,\sin(\varphi_{12}-\varphi_{23})
\right)
,
\end{eqnarray}
and the other components obtained by cP. This equation of motion is equivalent to (\ref{E0}) under the condition
\begin{equation}\label{E4}
{\mathbf S}\equiv{\mathbf r}_{12}+{\mathbf r}_{23}+{\mathbf r}_{31}={\mathbf 0}
\;.
\end{equation}
It can be shown that (\ref{E3a}) and (\ref{E3b}) imply  $\ddot{\mathbf S}={\mathbf 0}$, hence ${\mathbf S}$ must be a constant of motion
since it is a periodic function of time.
But it could assume any value. If the particular value ${\mathbf S}={\mathbf 0}$ is assumed for some solution of (\ref{E3a}) and (\ref{E3b}),
then the definitions
\begin{equation}\label{E4a}
{\mathbf r}_{1}\equiv \frac{1}{3}\left({\mathbf r}_{12}-{\mathbf r}_{31}\right)
\quad\mbox{ and cP for } {\mathbf r}_{2} \mbox{ and  } {\mathbf r}_{3}
\end{equation}
yields a solution of (\ref{E0}).

Next we make the FT ansatz
\begin{eqnarray}\label{E5a}
r_{12}(t)&=&\sum_{n=0}^\infty\,\sum_{m=n}^\infty\,\lambda^{2m}\,A_{2n,\,2m}\cos(4\,\pi\, n\,t),\\
\label{E5b}
r_{23}(t)&=&r_{12}(t-2\pi/3),\;r_{31}(t)=r_{12}(t+2\pi/3),\\
\label{E5c}
\varphi_{12}(t)&=&\lambda\,\cos(t)+\sum_{n=1}^\infty\,\sum_{m=n}^\infty\,\lambda^{2m+1}\,B_{2n+1,\,2m+1}\cos(2\,\pi\,(2 n+1)\,t)\\
\label{E5d}
\varphi_{23}(t)&=&r_{12}(t-2\pi/3),\;\varphi_{31}(t)=r_{12}(t+2\pi/3)
\;.
\end{eqnarray}
Thus $r_{12}$ is assumed to be an even $\cos$ series and $\varphi_{12}$ to be an odd one.
$\varphi_{ij}(t)$ could be written more generally to contain a non-zero mean value. This would be equivalent to a constant rotation of the
figure eight curve. We decided to set this mean value to zero and consequently obtain a figure eight curve standing upright, see below.
Moreover, $\lambda$ is ``concretized" by identifying it with the amplitude of the $\cos t$ mode of $\varphi_{12}$, i.~e.~,
\begin{equation}\label{E_lambda}
\lambda=\frac{1}{\pi}\,\int_0^{2\pi}\varphi_{12}(t)\,\cos t\; dt
\:.
\end{equation}
The numerical solution of \cite{CM00} corresponds to a value of
\begin{equation}\label{E_lambda_0}
\lambda=\lambda_0\approx 2.07568\;.
\end{equation}

It will be in order to justify the above FT ansatz by referring only to general properties of the figure eight solution, not to its
precise form according to the numerical solution obtainable from \cite{CM00}. This would support the claim that the FT series approximation is, in principle,
independent from the well-known numerical approximation.
We first observe that ${\mathbf r}_3(-t)=-{\mathbf r}_3(t)$, hence $x_3(t)$ and $y_3(t)$ can be expanded into $\sin$ series.
Anticipating that our ansatz leads to a figure eight curve standing upright, $x_3(t)$ must be an even $\sin$ series and $y(t)$ an odd one.
This follows from the symmetry properties at $t=\frac{T}{4}=\frac{\pi}{2}$, namely  $x_3(\pi/2-t)=-x_3(\pi/2+t)$ and  $y_3(\pi/2-t)=y_3(\pi/2+t)$.
By virtue of choreography, all $y_i(t)$ are odd Fourier series and all $x_i(t)$ even ones for $i=1,2,3$.

Since ${\mathbf r}_3(0)={\mathbf 0}={\mathbf r}_1(0)+{\mathbf r}_2(0)$, c.~f.~(\ref{E_CM1}), it follows by the symmetry of the figure eight solution
and (\ref{E_CM2}) that $\dot{\mathbf r}_1(0)=\dot{\mathbf r}_2(0)=-\frac{1}{2}\dot{\mathbf r}_3(0)$. Hence
${\mathbf r}_{12}(t)={\mathbf r}_{1}(t)-{\mathbf r}_{2}(t)$ has a stationary value at $t=0$, actually a local maximum. This makes it
(albeit not necessary but) highly plausible to choose $r_{12}(t)$ and  $\varphi_{12}(t)$  as $\cos$ series. Due to its definition
$r_{12}(t)$ will be an even $\cos$ series thereby justifying the ansatz (\ref{E5a}). Recall that all $y_i(t),\,i=1,2,3,$ and hence also $y_{12}(t)$ are
odd Fourier series. Since $y_{12}=r_{12}\,\sin\,\varphi_{12}$ the obvious way to guarantee this is to choose $\varphi_{12}(t)$ as an odd $\cos$ series.
This completes the justificaltion of (\ref{E5a}) and (\ref{E5c}).

Note further that
the choreographic ansatz in (\ref{E5b}) and (\ref{E5d}) can easily be evaluated by means of the addition theorem for the $\cos$ function.
This ansatz implies $r_{ij}=A_{00}+{\mathcal O}(\lambda),\;\varphi_{ij}={\mathcal O}(\lambda)$ and hence the r.~h.~s.~of (\ref{E3a}) and (\ref{E3a})
will be differentiable at $\lambda=0$. It turns out that this FT ansatz yields unique solutions for the coefficients $A_{nm}$ and $B_{nm}$
that can be calculated by computer algebraic means up to any desired order that is only limited by the available computer resources. \\
But we have got another problem: The condition (\ref{E4}) is in general not compatible
with (\ref{E5a}) to (\ref{E5d}). This can be easily seen by considering the limit $\lambda\rightarrow 0$, where all three
relative distances approach the value $A_{00}$ which means that the bodies form an equilateral triangle.
Hence one of the relative angles $\varphi_{ij}$ should
approach a non-zero value, for example $\frac{2\pi}{3}$ and not $0$. Another consequence of the violation of (\ref{E4}) is
that the angular momentum is no longer conserved.\\
Summarizing, we have obtained a $1$-parameter family of periodic solutions of (\ref{E3a}) and (\ref{E3b}),
but unfortunately of un-physical ones. Nevertheless, there is still a chance of analytically approximating the figure eight solution.
If the FT series is inserted, the quantity $\mathbf S$ becomes a function of $\lambda$. It can be shown that ${\mathbf S}_y(\lambda)$ always vanishes
since it is an odd Fourier series and a constant of motion.
On the other hand, ${\mathbf S}_x(\lambda)$ will only vanish for a discrete set of zeroes $\lambda_i$. For these values, the
FT series should give a physical solution of (\ref{E0}), especially the value $\lambda=\lambda_0$ should correspond
to the numerical solution of \cite{CM00}.
Hence we are left with the problem of determining the zeroes of ${\mathbf S}_x(\lambda)$ independently of the value $\lambda=\lambda_0$
obtained from the literature.
We will first illustrate the results of a low order truncation of the FT series.\\

The coefficients of the FT series up the $6$th order read
\begin{eqnarray}\label{E6a}
A_{00}&=& {3^{1/3}},\;A_{02}=-\frac{3^{1/3}}{16},\;A_{04}=\frac{3\times 3^{1/3}}{256},\\
\label{E6b}
A_{22}&=&\frac{3^{1/3}}{8},\;A_{24}=-\frac{3^{1/3}}{128},\;A_{26}=\frac{5\times 3^{1/3}}{4096},\\
\label{E6c}
A_{44}&=&\frac{3^{1/3}}{256},\;A_{46}=-\frac{29\times 3^{1/3}}{49152},\;A_{66}=-\frac{3^{1/3}}{8192},\\
\label{E6d}
B_{33}&=&-\frac{1}{48},\;B_{35}=-\frac{1}{512},\;B_{55}=-\frac{9}{20480}
\;.
\end{eqnarray}

\begin{figure}
\begin{center}
\includegraphics[clip=on,width=70mm,angle=0]{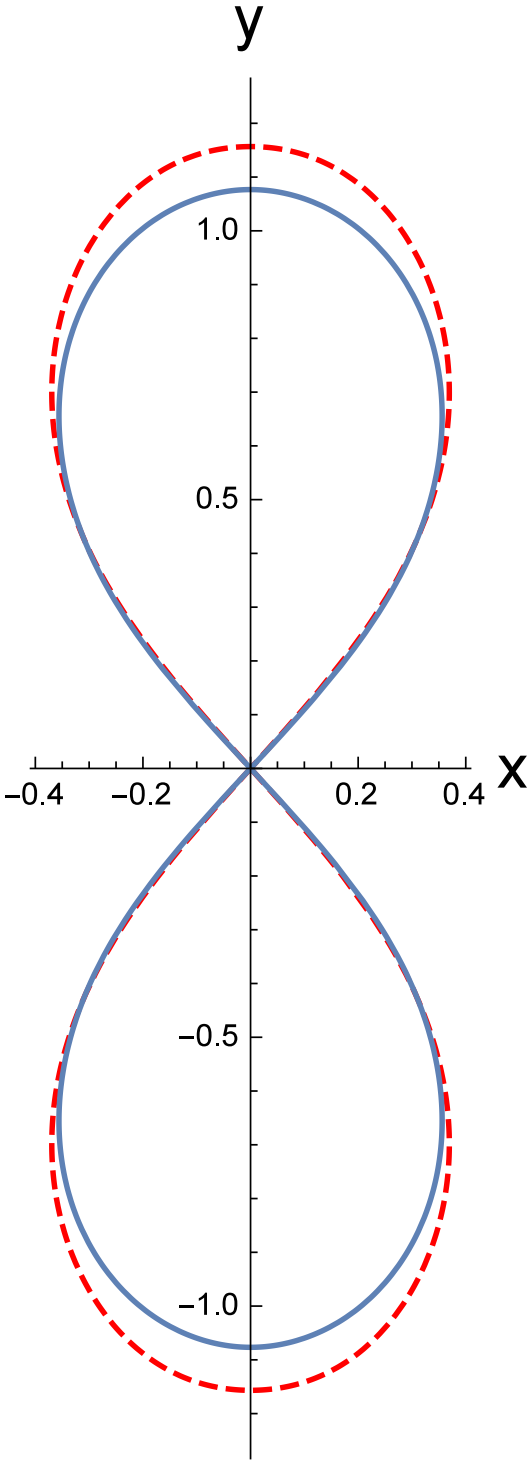}
\end{center}
\caption{Comparison of the figure eight solution curve corresponding to the FT series (\ref{E7a}),(\ref{E7b}) of order $6$
(red dashed curve)
with that of order $30$ (blue curve). The FT parameter $\lambda$ of the first curve is
chosen as $\lambda=\lambda_6\approx 2.09531$.
Note that in contrast to figure \ref{FIGE1} the figure eight is standing upright due to our choice
of the constant value for the FT series of $\varphi_{12}$ in (\ref{E5c}).
}
\label{FIGE4}
\end{figure}

This yields the following analytical approximations via (\ref{E4a}):
\begin{eqnarray}\label{E7a}
x_3(t)&\approx&
-\frac{\left(\lambda ^4-32 \lambda ^2+512\right) \lambda
   ^2 \sin (2 t)}{4096\; {3^{1/6}}}
-\frac{\left(\lambda ^2+192\right) \lambda ^4 \sin (4 t)}{49152\;
   {3^{1/6}}},\\
   \label{E7b}
y_3(t)&\approx&
-\frac{\lambda  \left(3
   \lambda ^4-32 \lambda ^2+256\right) \sin (t)}{256\; {3^{1/6}}}+
\frac{3\, \lambda ^5 \sin (5 t)}{4096\;{3^{1/6}}}
   \;.
\end{eqnarray}
This result seems reasonable since it correctly renders $x_3(t)$ as an even $\sin$ series
and $y_3(t)$ as an odd one. The coefficient of $\sin 3t$ must vanish since otherwise it would
lead to a non-zero value of $y_1(t)+y_2(t)+y_3(t)$ via choreography. In order to check the
degree of approximation we have plotted the orbit corresponding to (\ref{E7a}) and (\ref{E7b})
together with the orbit obtained by an FT series of order $30$ anticipating the result explained below.
For the first curve the FT parameter $\lambda$ is
chosen as the smallest real zero $\lambda_6\approx 2.09531$ of ${\mathbf S}(\lambda)$ evaluated up to $6$th order in $\lambda$.
The coincidence between both curves is not perfect, see figure \ref{FIGE4}, but the example demonstrates that
already an FT series approximation of low order gives a qualitatively correct picture of the figure eight solution.
By the way, the main source of the deviation is not the restriction to the order $6$ of the Fourier series
but the error of the Fourier coefficients due to the relatively large value of $\lambda_6$.\\

\begin{figure}
\begin{center}
\includegraphics[clip=on,width=100mm,angle=0]{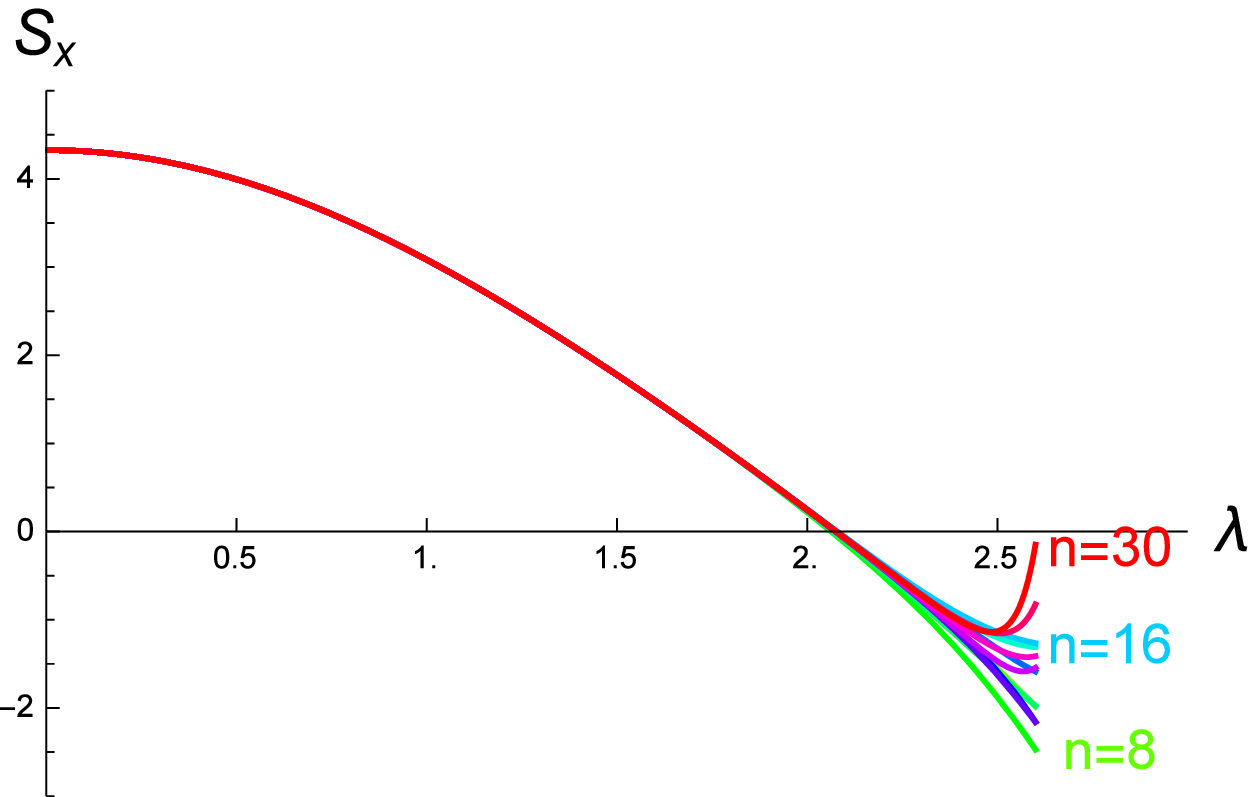}
\end{center}
\caption{Approximate graph of the function ${\mathbf S}_x(\lambda)$ defined in (\ref{E4}) corresponding to various orders $n=8,\ldots,30$
of truncation of the FT series. The vanishing of ${\mathbf S}_x(\lambda)$
is a necessary condition for the FT series becoming a physically meaningful approximation of the figure eight solution.
Accordingly we expect that the FT series parameter $\lambda$ should assume a value close to $2.07552$, see also figure \ref{FIGE7}.
}
\label{FIGE2}
\end{figure}

\begin{figure}
\begin{center}
\includegraphics[clip=on,width=100mm,angle=0]{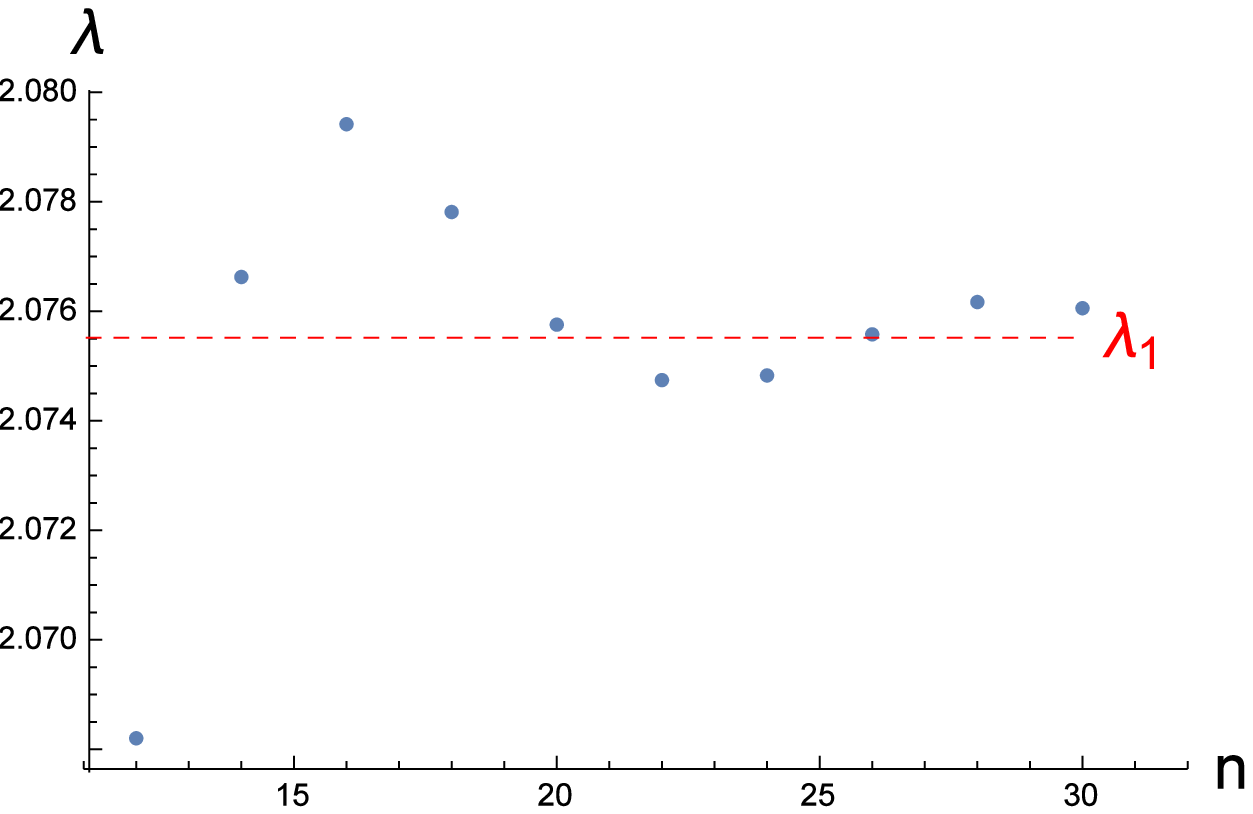}
\end{center}
\caption{Numerical zeroes $\lambda_n$ of ${\mathbf S}_x(\lambda)$ based on truncations of the FT series of order $n=12,\ldots,30$.
The mean value $\lambda_1=2.07552$ gives an estimate of the correct zero.}
\label{FIGE7}
\end{figure}

Next we extend the calculations to the $30$th order of the FT series. The detailed results for the series coefficients are too complicated
to be presented here and can be found in \cite{S15}. But we will show the graph of the function ${\mathbf S}_x(\lambda)$,
obtained by truncations of the FT series of order $n=8$ to $n=30$, see figure \ref{FIGE2}.
This graphs makes it plausible that ${\mathbf S}_x(\lambda)$ has a real zero close to $\lambda_0$, see (\ref{E_lambda_0}),
corresponding to the physical value of the series parameter,
but does not give a hint to the existence of further real zeroes. The zeroes $\lambda_n$ corresponding
to truncations of order $n$ can be numerically calculated and are shown in figure \ref{FIGE7}.
It seems that they perform a damped oscillation about their limiting value for $n\rightarrow\infty$.
Consequently, we take as an estimate of the true zero of ${\mathbf S}_x(\lambda)$ not $\lambda_0$
but the mean value of all $\lambda_n,\;n=12,14,\ldots,30$ that amounts to
\begin{equation}\label{E7}
\lambda_1 = 2.07552
\;,
\end{equation}
and has a relative deviation from the value $\lambda_0$ obtained
from \cite{CM00} of less than $8\times 10^{-5}$. Our method to
determine $\lambda_1$ could be criticized on grounds of its
arbitrariness and crudeness but we think that it would be pointless
to invoke more sophisticated methods that could only slightly
improve the deviation from the ``correct" value of the zero.

\begin{figure}
\begin{center}
\includegraphics[clip=on,width=100mm,angle=0]{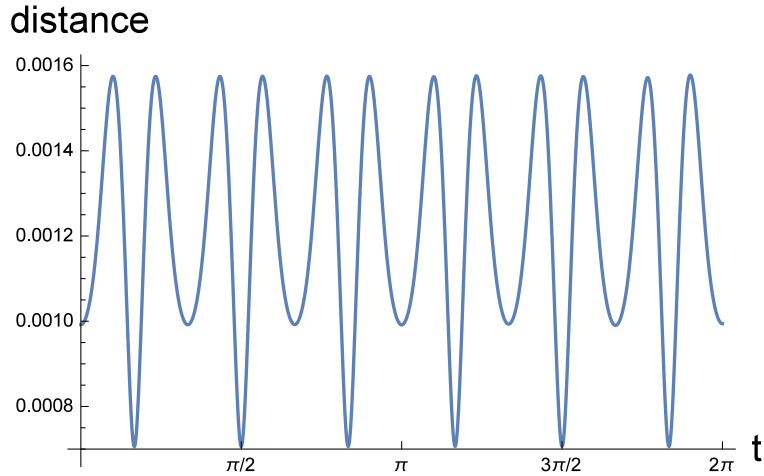}
\end{center}
\caption{Plot of the Euclidean distance as a function of time $t$ between the analytical approximation based on an FT series of order $30$
and the numerical figure eight solution based on the initial values published in \cite{CM00}.
}
\label{FIGE6}
\end{figure}

The next task is to calculate the figure eight solution corresponding to the calculated series coefficients
of order $30$ and the estimate $\lambda_1$ of the physical value of the FT series parameter $\lambda$.
The result for the orbit has already be presented in figure \ref{FIGE4}. The difference to the figure eight
curve obtained from the data of \cite{CM00} would not be visible. Hence we have plotted the Euclidean
distance $\left(\sum_{i=1}^3 |{\mathbf r}_i^{FT}(t)-{\mathbf r}_i^{CM}(t)|^2\right)^{1/2}$ as a function of $t$,
see figure \ref{FIGE6},
where ${\mathbf r}_i^{FT}(t)$ denotes the analytical approximation based on the FT series of order $30$
and ${\mathbf r}_i^{CM}(t)$ the numerical figure eight solution according to the initial values published in \cite{CM00}.
It is clear that we had to modify these initial values by a scale (``Kepler") transformation
and by suitable reflections for the sake of comparison with the FT series approximation.
The result is a maximal deviation of $1.6\times 10^{-3}$ between both solutions that is larger than
expected but explains why the difference between both figure eight curves would not be visible.

\section{Summary and Outlook}
\label{sec:S}
We have proposed a method suited to analytically approximate periodic solutions of non-linear equations of motion,
the Fourier Taylor (FT) series. If the method works it yields the series coefficients in a recursive way
such that each step involves a new Fourier order and simultaneously improves the accuracy of the old
Fourier coefficients. Typical applications of this method include mechanical oscillations about some ground state
and extend the validity of the approximation beyond the linear regime, as in the example of section \ref{sec:C}.
Even for linear problems the FT method can be used as a simpler alternative to perturbation theory, see section \ref{sec:M}.
In these cases the FT series yields a family of solutions depending on a parameter $\lambda$
and the domain of application is limited by the convergence radius w.~r.~t.~$\lambda$ and, additionally,
by the possible truncation order of the FT series. In most cases the FT coefficients can only be calculated
by the aid of computer algebra and hence the maximal truncation order is given by the available
computer sources. Nevertheless, the FT series approximation has some advantages compared with
a purely numerical treatment of the problem: For example, the FT coefficients may depend on further
parameters of the problem and thus an FT approximation may be equivalent to an infinite number
of numerical calculations for different values of the parameter, just as in the case where
a closed formula for a class of solutions has been found.\\
From this point of view the application of the FT series method to the figure eight problem
is somewhat untypical. We could not directly apply the method to the equation of motion
but only to another equation for the relative coordinates that follows from the original one.
Conversely, the new equation implies the original one only if some addition condition (\ref{E4}) is satisfied.
The FT series violates this condition, and thus produces un-physical solutions, except for
a particular value $\lambda_1$ of the series parameter $\lambda$. For this value the FT series
approximates the figure eight solution known from \cite{CM00}, where the error depends on the truncation order.
In principle it can be arbitrarily small, in practice we have to be content with an error of $1.6\times 10^{-3}$
for a truncation order of $n=30$.\\
One may ask: What is the particular virtue of the FT series in the figure eight example,
except for illustrating the scope of the method? There exist already two ways to access the problem,
namely the numerical solution found by some educated guess \cite{M93}, \cite{CM00} and the mathematical existence proof \cite{CM00}.
Compared with both ways the FT series method seems to be inferior: The result obtainable in practice is less accurate than the numerical one
and it is not rigorous in so far as the convergence of the FT series has not been proven.
We think, however, that the virtue of the FT series method is, first, that it provides a
third approach to the figure eight problem that is, in principle, independent of the two other ones.
For this reason we have carefully identified the assumptions leading to the FT ansatz in section \ref{sec:E}.
Secondly, the FT series method requires less ingenuity, so to speak, compared with the two other methods.
Once the adequate ansatz has been found, the method works automatically and can (and should) be performed
by computer software. Hence in this way it might be possible to easier find new solutions of the three body problem that
are conjectured to exist and to have special symmetries. But this is a task for future work and beyond the scope of the present paper.

\section*{References}


\begin{thebibliography}{99}
\bibitem{S15}Three MATHEMATICA notebooks containing applications of the FT series method are available at
\verb" msuq.physik.uni-osnabrueck.de/tools.htm"
\end{thebibliography}
\end{document}